\documentstyle[preprint,epsf]{jpsj}

\def\runtitle{
A Quantum Monte Carlo Method and Its Applications
to Multi-Orbital Hubbard Models
}
\def\runauthor{Yukitoshi {\sc Motome} and Masatoshi {\sc Imada}}

\def\sf{\rm}

\title{
A Quantum Monte Carlo Method and Its Applications\\
to Multi-Orbital Hubbard Models
}

\author{Yukitoshi {\sc Motome} and Masatoshi {\sc Imada}
}
\inst{
Institute for Solid State Physics, University of Tokyo, \\
Roppongi 7-22-1, Minato-ku, Tokyo 106
}
\recdate{
April 24, 1997
}

\abst{
We present a framework of 
an auxiliary field quantum Monte Carlo (QMC) method
for multi-orbital Hubbard models.
Our formulation can be applied to 
a Hamiltonian which includes terms for 
on-site Coulomb interaction for both intra- and inter-orbitals,
intra-site exchange interaction and 
energy differences between orbitals.
Based on our framework,
we point out possible ways
to investigate various phase transitions
such as metal-insulator, magnetic and orbital order-disorder transitions
{\it without the minus sign problem}.
As an application,
a two-band model is investigated 
by the projection QMC method and
the ground state properties of this model are presented.
}

\kword{
auxiliary field quantum Monte Carlo,
multi-orbital Hubbard model,
minus sign problem,
metal-insulator transition,
magnetic ordering,
orbital ordering
}

\begin{document}
\sloppy
\maketitle

Recently,
much attention has been directed to
the relevance of orbital degrees of freedom
in strongly correlated electron systems.
In addition to the coupling to the lattice distortion,
fluctuations of the orbital component
may cause fruitful interplay
with those of spin degrees of freedom.
This competition together with the strong electron correlation
results in a variety of remarkable physical properties,
for example,
metal-insulator transition,
spin and orbital order-disorder transitions
and their fluctuations.

In spite of their relevance to many aspects
of physical properties in $d$ and $f$ electron compounds,
theoretical as well as experimental approaches 
to these complex phenomena
have not been fully developed so far.
In particular,
as compared with relatively simple and symmetric couplings
in the spin degrees of freedom,
the orbital degeneracy includes rather complicated elements
which should be taken into account in theoretical models.

One of the minimal models for this approach is
the multi-orbital Hubbard model.
Numerous investigations based on various approximations
have been performed on this model.
\cite{Roth1,Roth2,Cyrot,Inagaki1,Inagaki2,Kugel1,Kugel2,Chao,Lu,Okabe,Bunemann,Kotliar,Rozenberg1,Rozenberg2,Ishihara,Fresard,Hasegawa1,Hasegawa2}
However,
for study of the serious competition mentioned above and
discussion of the critical properties
of these phase transitions in this system,
a more strict treatment beyond these approximations is desired.

In this letter,
we propose a framework of 
an auxiliary field QMC technique
for this type of multi-orbital Hubbard models.
Our technique makes it possible to study
systems with any orbital degeneracies in any dimensions.
Based on our algorithm,
we identify some parameter regions
in which no minus sign problem occurs.
Using these parameter sets,
we point out that
detailed investigations
on various phase transitions,
such as metal-insulator transitions,
magnetic and orbital orderings become tractable.
In order to demonstrate the efficiency of our framework,
we use it to investigate a two-band model
containing level splitting
and next-nearest neighbor hopping.
Using the projection QMC method
completely free from the minus sign problem,
we discuss ground state properties of this two-band model.

Here, we consider the multi-orbital Hubbard Hamiltonian.
The explicit form of our Hamiltonian is given by
\begin{equation}
\label{DHMHamiltonian}
{\cal H} = {\cal H}_{t} + {\cal H}_{U} 
+ {\cal H}_{J} + {\cal H}_{\varepsilon}, \\
\end{equation}
where
\begin{eqnarray}
{\cal H}_{t} &=& \sum_{i,j} \sum_{\nu,\nu'} \sum_{\sigma}
                          t_{ij}^{\nu\nu'} 
                          \bigl( c_{i\nu\sigma}^{\dagger} c_{j\nu'\sigma}
                          + \mbox{h.c.} \bigr) \\
{\cal H}_{U} &=& \sum_{i} \sum_{\nu\leq\nu'} \sum_{\sigma\leq\sigma'}
                          \left(1-\delta_{\nu\nu'}\delta_{\sigma\sigma'}\right)
                          U_{\nu\nu'} n_{i\nu\sigma} n_{i\nu'\sigma'}
\label{HU} \\
{\cal H}_{J} &=& - \sum_{i} \sum_{\nu<\nu'} \sum_{\sigma\sigma'}
                          J_{\nu\nu'}
                          \bigl( c_{i\nu\sigma}^{\dagger} c_{i\nu\sigma'}
                                 c_{i\nu'\sigma'}^{\dagger} c_{i\nu'\sigma}
\nonumber \\
             & &   \qquad \qquad \qquad \qquad
                               + c_{i\nu\sigma}^{\dagger} c_{i\nu'\sigma'}
                                 c_{i\nu\sigma'}^{\dagger} c_{i\nu'\sigma}
                          \bigr)
\label{HJ} \\
{\cal H}_{\varepsilon} &=& \sum_{i} \sum_{\nu} \varepsilon_{\nu} n_{i\nu}.
\end{eqnarray}
This model has $N_{S}$ sites on a bipartite lattice
and $N_{D}$ orbitals per site.
The operator $c_{i\nu\sigma}^{\dagger}$ creates
the $\sigma$-spin electron at site $i=1,\cdot\cdot\cdot,N_{S}$
and in orbital $\nu=1,\cdot\cdot\cdot,N_{D}$.
Here, 
$t_{ij}^{\nu\nu'}$ is the hopping integral
between orbital $\nu$ at site $i$ and
orbital $\nu'$ at site $j$;
$U_{\nu\nu'}$ is the on-site Coulomb interaction 
between orbitals $\nu$ and $\nu'$;
$J_{\nu\nu'}$ is the intra-site exchange interaction;
and $\varepsilon_{\nu}$ is the level energy for orbital $\nu$.

Two-body interactions $U_{\nu\nu'}$ and $J_{\nu\nu'}$ are
not originally independent.
There is a relation
$U_{\nu\nu} = U_{\nu\ne\nu'} - 2J_{\nu\nu'}$,
because of the rotational symmetry of the Coulomb terms.
\cite{Kanamori,Brandow}
However, in our model,
we treat them as independent parameters.
Moreover,
we take $U_{\nu\nu}=U_{\nu\neq\nu'} \equiv U \geq 0$ and
$J_{\nu\nu'} \geq 0$
for the convenience of our QMC algorithm.
Then, we can factorize the two-body interaction terms
(\ref{HU}) and (\ref{HJ})
into the quadratic forms, as
\begin{eqnarray}
{\cal H}_{U} &=& \frac{U}{2} \sum_{i} \left(n_{i}-\frac{N_{D}}{2}\right)^{2}
+ \frac{U}{2} \Bigl(N_{D}-1\Bigr) \sum_{i} n_{i}\label{HU2} \\
{\cal H}_{J} &=& \sum_{i} \sum_{\nu<\nu'} 
                 \frac{J_{\nu\nu'}}{2} A_{i\nu\nu'}^{2} \nonumber \\
             &-& \Bigl(N_{D}-1\Bigr) \sum_{i} \sum_{\nu<\nu'} 
                 \frac{J_{\nu\nu'}}{2} n_{i}.
\label{HJ2}
\end{eqnarray}
Here, 
$A_{i\nu\nu'} \equiv \sum_{\sigma} A_{i\nu\nu'\sigma}
\equiv \sum_{\sigma} 
( c_{i\nu\sigma}^{\dagger} c_{i\nu'\sigma}
+ c_{i\nu'\sigma}^{\dagger} c_{i\nu\sigma} )$.

Now, we develop the auxiliary field QMC algorithm
for model (\ref{DHMHamiltonian}).
The prescription explained below is for the ground-state
which is called the projection QMC algorithm.\cite{White,Imada}
It may be straightforwardly extended to
the finite-temperature algorithm.\cite{Hirsch,Sugiyama}

In the ground-state algorithm,
we need to calculate the density matrix,
$\rho(\tau;\phi)=\langle\phi|\exp(-\tau{\cal H})|\phi\rangle$,
where $|\phi\rangle$ is a trial wave function
non-orthogonal to the ground state.
After the Suzuki-Trotter decomposition
of $\exp(-\tau{\cal H})$ into the $M=\tau/\Delta\tau$ slices,
we replace the two-body interaction terms
with non-interacting ones by introducing
summations over the Hubbard-Stratonovich variables.
The general formula of the discrete Hubbard-Stratonovich transformation
we use here is given by
\begin{eqnarray}
& & \exp\left(-\Delta\tau \theta f^{2}\right) \nonumber \\
&=& \sum_{l=\pm1} \sum_{s=\pm1} \frac{\gamma_{l}}{4}
\exp\left( i s \eta_{l} \sqrt{\alpha} f \right) + O(\Delta\tau^{4}),
\end{eqnarray}
where
$\gamma_{l} \equiv 1+\frac{\sqrt{6}}{3}l,
\eta_{l} \equiv \sqrt{2(3-\sqrt{6}l)}$
and $\alpha \equiv \Delta\tau \theta \geq 0$.
Using this formula,
we decouple the two-body interaction terms in
(\ref{HU2}) and (\ref{HJ2}), as
\begin{eqnarray}
& & \exp \Bigl( -\Delta\tau 
\frac{U}{2} \sum_{i} \left( n_{i}-\frac{N_{D}}{2}\right)^{2} \Bigr) 
\nonumber \\
&=& \prod_{i} \biggl[ \sum_{l_{1}=\pm1} \sum_{s_{1}=\pm1}
\frac{\gamma_{l_{1}}}{4} \exp \left\{ i s_{1} \eta_{l_{1}}
\sqrt{\alpha_{1}} \left( n_{i}-\frac{N_{D}}{2} \right) \right\} \biggr] 
\nonumber \\
& & \qquad \qquad \qquad \qquad \qquad \qquad \qquad \quad 
+ O(\Delta\tau^{4})
\label{expU} \\
& & \exp \Bigl( -\Delta\tau 
\sum_{i}\sum_{\nu<\nu'} \frac{J_{\nu\nu'}}{2} 
A_{i\nu\nu'}^{2} \Bigr) \nonumber \\
&=& \prod_{i} \prod_{\nu<\nu'} \biggl[ \sum_{l_{2}=\pm1} \sum_{s_{2}=\pm1}
\frac{\gamma_{l_{2}}}{4} \exp \left\{ i s_{2} \eta_{l_{2}}
\sqrt{\alpha_{2}} A_{i\nu\nu'} \right\} \biggr] \nonumber \\
& & \qquad \qquad \qquad \qquad \qquad \qquad \qquad \quad 
+ O(\Delta\tau^{4}),
\label{expJ}
\end{eqnarray}
where $\alpha_{1}=\Delta\tau U/2$ and $\alpha_{2}=\Delta\tau J_{\nu\nu'}/2$.

We apply these decompositions to each Suzuki-Trotter slice.
Then, the one-body expression is given as
the product of these terms
together with the non-interacting terms 
of the Hamiltonian (\ref{DHMHamiltonian}), as
\begin{eqnarray}
\rho\left(\tau;\phi\right) =
\sum_{\{l_{1}s_{1}l_{2}s_{2}\}}
& & W_{\uparrow}\left(\{l_{1}s_{1}l_{2}s_{2}\};\tau;\phi\right)
\nonumber \\
& & W_{\downarrow}\left(\{l_{1}s_{1}l_{2}s_{2}\};\tau;\phi\right)
\label{densitymat}
\end{eqnarray}
where $W_{\sigma}$ is given by
\begin{eqnarray}
W_{\sigma} =
\langle\phi_{\sigma}| \prod_{m=1}^{M} & & \bigl[
w_{t\sigma} w_{U\sigma}(l_{1}(m),s_{1}(m))
\nonumber \\
& & w_{J\sigma}(l_{2}(m),s_{2}(m)) 
w_{\varepsilon\sigma} w_{t\sigma} \bigr] |\phi_{\sigma}\rangle
\end{eqnarray}
with
\begin{eqnarray}
w_{t\sigma} &=& \prod_{i=1}^{N_{S}} 
\exp\left(-\Delta\tau{\cal H}_{t\sigma}/2\right) \\
w_{U\sigma} &=& \prod_{i=1}^{N_{S}}
\left[ \frac{\sqrt{\gamma_{l_{1i}}}}{2}
\exp \Bigl\{ is_{1i}\eta_{l_{1i}} \right. \nonumber \\
& & \left. \qquad\qquad\quad\quad\quad
\sqrt{\alpha_{1}} \Bigl(n_{i\sigma}-\frac{N_{D}}{2}
\Bigr) \Bigr\} \right] \\
w_{J\sigma} &=& \prod_{i=1}^{N_{S}} \prod_{\nu<\nu'}
\Bigl[ \frac{\sqrt{\gamma_{l_{2i}}}}{2}
\exp \left\{ is_{2i}\eta_{l_{2i}}
\sqrt{\alpha_{2}} A_{i\nu\nu'\sigma} \right\} \Bigr] \\
w_{\varepsilon\sigma} &=& \prod_{i=1}^{N_{S}} \prod_{\nu}
\exp\left( -\Delta\tau \varepsilon_{\nu} n_{i\nu\sigma} \right).
\end{eqnarray}
The product of eqs. (\ref{expU}) and (\ref{expJ})
over all slices amounts to 
the systematic error of $O(\Delta\tau^{3})$
which is negligible because it is higher order than
the other systematic errors 
from the Suzuki-Trotter decomposition.

The summations over $\{l_{1}s_{1}l_{2}s_{2}\}$ 
in eq. (\ref{densitymat}) are
statistically sampled by the Monte Carlo technique.
The product $W_{\uparrow} W_{\downarrow}$ plays the role of
the weight for each sample in the QMC updates.
In general,
this weight can have a negative value,
which leads to
the minus sign problem in the QMC calculation.
The difficulty caused by the minus sign
depends on the electron density, the long-range hopping
and the dimensionality of the system.
The details of this problem for this model
will be reported elsewhere.

However, the above formulation provides us
with a useful property in the QMC updates.
That is, 
the minus sign problem does not appear
for some regions of parameter values in our Hamiltonian.
For the parameters explained below,
it is easily shown that
the weight of the QMC sample for the up-spin is
simply the complex conjugate of that for the down-spin;
$W_{\uparrow} W_{\downarrow} = |W_{\uparrow}|^{2} \ge 0$.

The simplest case of the parameter sets
which are free from the minus sign problem
are given by the following conditions:
(i) The system is at half filling; $n_{e}=N_{D}$.
(ii) All orbitals are completely degenerated; $\varepsilon_{\nu}=0$.
(iii) The hopping integrals are non-zero 
only between the nearest neighbor sites.
Under these constraints,
we can easily show the complex conjugate relation
between the QMC weights for the up- and down-spin
by the particle-hole transformation
\begin{eqnarray}
c_{i\nu\uparrow} &\rightarrow& 
(-1)^{i} {\tilde c}_{i\nu\uparrow}^{\dagger} \\
c_{i\nu\downarrow} &\rightarrow& 
{\tilde c}_{i\nu\downarrow}.
\end{eqnarray}
This is a straightforward extension
of that for the one-band Hubbard model at half filling.

Using the symmetry about the orbital indices,
the above conditions for the absence of the minus sign problem
may easily be relaxed and generalized to the following nontrivial ones:
(i) The system is at half filling; $n_{e}=N_{D}$.
(ii) The energy levels split symmetrically
around the Fermi energy;
$\varepsilon_{\nu} = - \varepsilon_{\bar{\nu}}$,
where ${\bar \nu} \equiv N_{D}-\nu+1$.
(iii) The hopping integrals satisfy a special condition;
$t_{ij}^{\nu\nu'} = (-1)^{|i-j|} t_{ij}^{{\bar {\nu'}}{\bar \nu}}$,
where $|i-j|$ is the Manhattan distance between site $i$ and $j$.
(iv) The intra-site exchange couplings also satisfy a special relation;
$J_{\nu\nu'} = \delta_{{\bar \nu}\nu'} J_{\nu}$.
For these sets,
the particle-hole transformation
to confirm the positivity of the QMC weights is
\begin{eqnarray}
c_{i\nu\uparrow} &\rightarrow& 
(-1)^{i} {\tilde c}_{i{\bar \nu}\uparrow}^{\dagger} \\
c_{i\nu\downarrow} &\rightarrow& 
{\tilde c}_{i\nu\downarrow}.
\end{eqnarray}

Within the former parameter sets,
the perfect nesting property
of the non-interacting energy dispersion remains unchanged,
although we can change the bandwidths of each orbital independently.
Therefore,
the system is expected to become the Mott insulator
for any finite values of $U$ and $J$.
Compared with this, however,
the latter constraints have much more generality with an advantage
for our purpose to investigate various phase transitions
in this complex system.
Both conditions (ii) and (iii) offer us
the possibility to investigate metal-insulator transitions,
since each of them breaks the perfect nesting property
in different ways.
These transitions are also expected to trigger
order-disorder transitions of the spin and orbital components.

To sum up our statement:
Our QMC framework explained above
provides ways of investigating various phase transitions
in an $N_{D}$-orbital Hubbard model 
(\ref{DHMHamiltonian}) at half filling
{\it completely free from the minus sign problem}.
Control parameters are
the Coulomb interaction $U$,
the intra-site exchange couplings $J_{\nu\nu'}$,
the level splitting $\varepsilon_{\nu}$ and 
the hopping integrals $t_{ij}^{\nu\nu'}$
with some constraints on the latter two
to keep the particle-hole symmetry.
These wide parameter choices allow us
to study the metal-insulator transition,
the spin and orbital ordering transitions.

In the following,
we present an example of the applications.
Here, we consider a two-band model
with the latter choice of the above constraints;
(i) $n_{e}=2$,
(ii) $\varepsilon_{\nu=1} 
= - \varepsilon_{\nu=2} \equiv \varepsilon$ and
(iii) $t_{ij}^{\nu\nu'} = -\delta_{\nu\nu'} t$ 
for the nearest neighbor sites
and $t_{ij}^{\nu\nu'} = -\delta_{\nu\nu'} (-1)^{\nu} t'$
for the next-nearest neighbor sites.

This model has two trivial limits:
(a) When $U \gg J \gg t$ and $t'$,
the system is the Mott insulator.
In this limit, the Hamiltonian is rewritten as
the $S=1$ spin system by the second-order perturbation.
(b) When $\varepsilon \gg U,J,t$ and $t'$,
since all the electrons fully occupy the orbital $\nu=2$,
the system becomes a band insulator.

For finite values of $U$ and $J$,
the system is expected to be in the Mott insulating state
at $\varepsilon=t'=0$
because of the perfect nesting.
From a consideration for the weak correlation limit,
a finite $\varepsilon$ causes level splitting and
a finite $t'$ results in 
deformation of the Fermi surfaces of each orbital.
Both should lead to self-doping between two bands.
In the Hartree-Fock calculation,
we find that
these effects cause metal-insulator transitions.
Depending on the choice of the route of the transitions,
magnetic or orbital orderings
take place either accompanied by the metal-insulator transition
or independently.
Therefore,
the rich interplay of these transitions deserves study
in a more reliable way.

We present here preliminary results for
the ground state properties of this model in one dimension (1D)
calculated by the projection QMC technique
to demonstrate the efficiency of our method.
In all the calculations,
we fix $U/t=2$ and $J/t=0.5$.
We set $\Delta\tau=t/20$ which ensures within the statistical errors
the convergence to the limit $\Delta\tau \rightarrow 0$
for all physical quantities we calculated.
The following data are obtained from $2000\sim6000$ QMC averages
for the state $\exp(-\frac{\tau}{2}{\cal H})|\phi\rangle$,
where we take $\tau=10\sim40t$ for each situation
to obtain converged values in the ground state.
The unrestricted Hartree-Fock ground state is used for $|\phi\rangle$
to accelerate the convergence.\cite{Furukawa}

In the following,
we focus on the correlation functions
for the spin and iso-spin degrees of freedom;
\begin{equation}
O^{\alpha} \left({\vec q}\right) =
\frac{1}{N_{S}} \sum_{i,j} 
\langle O_{i}^{\alpha} \cdot O_{j}^{\alpha} \rangle 
\exp\left(i{\vec q}\cdot{\vec r}_{ij}\right), \\
\end{equation}
where $\alpha=x,y,z$.
$O=S$ and $T$ are the spin and iso-spin operators, defined as
\begin{eqnarray}
S_{i}^{\alpha} &=& \frac{1}{2}
\sum_{\nu} \sum_{\sigma\sigma'}
\vec{\tau}_{\sigma\sigma'}^{\alpha}
c_{i\nu\sigma}^{\dagger} c_{i\nu\sigma'} \\
T_{i}^{\alpha} &=& \frac{1}{2}
\sum_{\sigma} \sum_{\nu\nu'}
\vec{\tau}_{\nu\nu'}^{\alpha}
c_{i\nu\sigma}^{\dagger} c_{i\nu'\sigma},
\end{eqnarray}
where $\vec{\tau}$ is the Pauli matrix.
Since the Hamiltonian (\ref{DHMHamiltonian}) has
rotational symmetry about the spin,
we calculate the spin correlation using
$S = (S^{x}+S^{y}+S^{z})/3$.
However,
since the iso-spin has an easy axis in the $z$-direction,
we focus on $T^{z}$.

\begin{figure}
\epsfbox{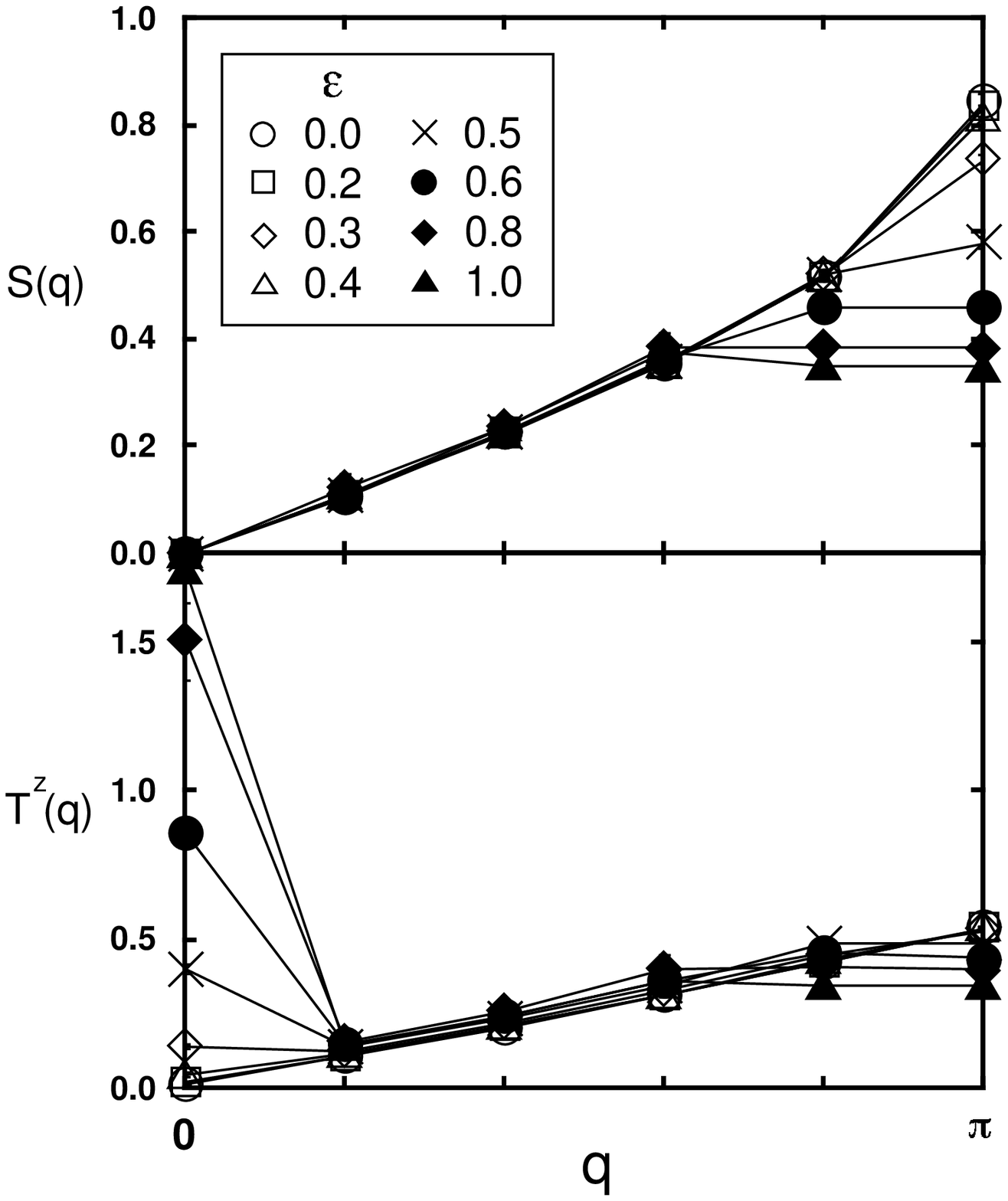}
\caption{
QMC results for the spin (upper-half) and 
iso-spin (lower-half) correlation functions at $t'=0$
for the two-band model in one dimension (10 sites).
Each symbol denotes the difference of $\varepsilon$ in units of $t$
as indicated in the inset.
}
\label{corr:t'=0}
\end{figure}

First,
we show typical $\varepsilon$ dependence of 
two correlation functions at $t'=0$
in Fig. \ref{corr:t'=0}.
For small values of $\varepsilon$,
the spin correlation is enhanced at $q=\pi$ and
the iso-spin correlation function has no characteristic structure
compared with the non-interacting case.
When $\varepsilon$ increases,
the enhancement of $S(q=\pi)$ disappears and
the $q=0$ component of the iso-spin correlation grows markedly.
Moreover,
both correlation functions have cusp-like structures at $q=2k_{F}$,
where $k_{F}$ is the Fermi wave number of 
the non-interacting counterparts of these systems.
This suggests that
the level splitting leads to
the self-doping from the orbital $\nu=1$ to $2$ accompanied by
the collapse of the nesting property away from $\varepsilon=0$.

\begin{figure}
\epsfbox{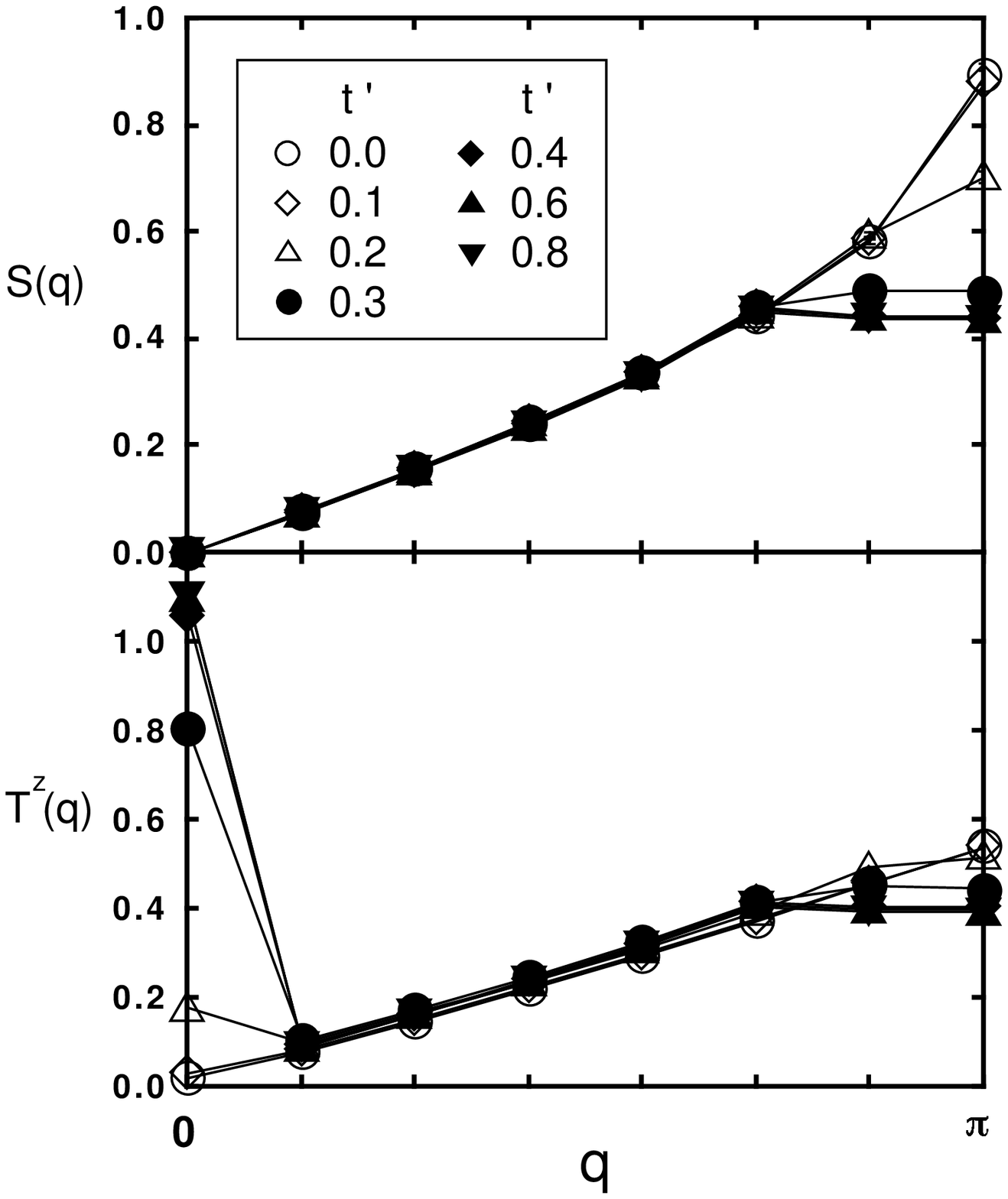}
\caption{
QMC results for the spin (upper-half) and 
iso-spin (lower-half) correlation functions at $\varepsilon=0$
for the two-band model in one dimension (14 sites).
Each symbol denotes the difference of $t'$ in units of $t$
as indicated in the inset.
}
\label{corr:Dlv=0}
\end{figure}

Next,
we show the dependence on $t'$ at $\varepsilon=0$
in Fig. \ref{corr:Dlv=0}.
As in Fig. \ref{corr:t'=0},
they also suggest a drastic change
due to the self-doping between two orbitals.

These QMC data clearly show the self-doping effects
by the level-splitting or the next-nearest neighbor hopping
as expected in this model.
Moreover, the momentum distribution function clearly shows
an evidence for the self-doping.
We can calculate the electron density in each orbital
by integrating the momentum distribution function for each orbital,
and find the difference of densities between two orbitals
for large values of $\varepsilon$ and $t'$.
A more detailed study and analyses 
on the nature of these substantial changes
in 1D as well as in 2D
will be reported elsewhere.

To summarize,
we have developed a new framework of 
an auxiliary field QMC technique
for multi-orbital Hubbard models.
Our Hamiltonian includes terms for
intra- and inter-orbital Coulomb interaction,
intra-site exchange interaction and 
level differences between orbitals.
Systems with some parameter sets are completely
free from the minus sign problem.
For these parameter sets,
we have pointed out the possibility
to investigate various phase transitions.
An example of the two-band model has been studied, and
the ground state properties of this model are calculated
based on our algorithm.
The QMC data in one dimension clearly show
the substantial changes
induced by the self-doping between two orbitals.

The authors thank N. Furukawa for fruitful discussions.
This work is supported by a Grant-in-Aid
for Scientific Research on the Priority Area
`Anomalous Metallic State near the Mott Transition'
from the Ministry of Education, Science, Culture an Sports, Japan.
The computations in this work were performed
using the facilities of the Supercomputer Center,
Institute for Solid State Physics, University of Tokyo.

\begin{thebibliography}{99}

\bibitem{Roth1} L. M. Roth: Phys. Rev. {\bf 149} (1966) 306.
\bibitem{Roth2} L. M. Roth: J. Appl. Phys. {\bf 38} (1967) 1065.
\bibitem{Cyrot} M. Cyrot and C. Lyon-Caen: J. Phys. {\bf 36} (1975) 253.
\bibitem{Inagaki1} S. Inagaki and R. Kubo: Int. J. Mag. {\bf 4} (1973) 139.
\bibitem{Inagaki2} S. Inagaki: J. Phys. Soc. Jpn. {\bf 39} (1975) 596.
\bibitem{Kugel1} K. I. Kugel' and D. I. Khomski\u\i: 
Sov. Phys. Ups. {\bf 25} (1982) 231.
\bibitem{Kugel2} K. I. Kugel' and D. I. Khomski\u\i:
Sov. Phys. JETP. {\bf 37} (1973) 725.

\bibitem{Chao} K. A. Chao and M. C. Gutzwiller:
J. Appl. Phys. {\bf 42} (1971) 1420.
\bibitem{Lu} J. P. Lu: Phys. Rev. B {\bf 49} (1994) 5687.
\bibitem{Okabe} T. Okabe: J. Phys. Soc. Jpn. {\bf 65} (1996) 1056.
\bibitem{Bunemann} J. B\"unemann and W. Weber: preprint(cond-mat/9611032).

\bibitem{Kotliar} G. Kotliar and H. Kajueter:
Phys. Rev. B {\bf 54} (1996) 14221.
\bibitem{Rozenberg1} M. J. Rozenberg: preprint(cond-mat/9611045).
\bibitem{Rozenberg2} M. J. Rozenberg: preprint(cond-mat/9612089).

\bibitem{Ishihara} S. Ishihara, M. Yamanaka and N. Nagaosa:
preprint(cond-mat/9606160).
\bibitem{Fresard} R. Fr\'esard and G. Kotliar: preprint(cond-mat/9612172).
\bibitem{Hasegawa1} H. Hasegawa: preprint(cond-mat/9612142).
\bibitem{Hasegawa2} H. Hasegawa: preprint(cond-mat/9702034).


\bibitem{Kanamori} J. Kanamori: Prog. Theor. Phys. {\bf 30} (1963) 275.
\bibitem{Brandow} B. H. Brandow: Adv. Phys. {\bf 26} (1977) 651.


\bibitem{White} S. R. White, D. J. Scalapino, R. L. Sugar,
E. Y. Loh, J. E. Gubernatis and R. T. Scalettar:
Phys. Rev. B {\bf 40} (1989) 506.
\bibitem{Imada} M. Imada and Y. Hatsugai:
J. Phys. Soc. Jpn. {\bf 58} (1989) 3752.

\bibitem{Hirsch} J. E. Hirsch: Phys. Rev. B {\bf 28} (1983) 4059.
\bibitem{Sugiyama} G. Sugiyama and S. E. Koonin:
Annals of Phys. {\bf 168} (1986) 1.

\bibitem{Furukawa} N. Furukawa and M. Imada:
J. Phys. Soc. Jpn. {\bf 60} (1991) 3669.

\end{thebibliography}
\end{document}